\documentclass[twocolumn]{aastex631}

\usepackage{graphicx}	
\usepackage{amsmath}	
\usepackage{amssymb}	
\usepackage{comment}
\usepackage{braket}
\usepackage{booktabs}
\usepackage{tikz}

\newcommand\HI{H{\small{I}}}
\received{\today}
\revised{YYY}
\accepted{ZZZ}

\shorttitle{Evolution of \HI{} galaxy scaling relations with stacking at $z<0.5$}
\shortauthors{Sinigaglia et al.}
\graphicspath{{./}{figures/}}

\begin{document}

\title{MIGHTEE-H{\small I}: Evolution of H{\small I} scaling relations of star-forming galaxies at $z<0.5$\footnote{Released on \today}}

\correspondingauthor{Francesco Sinigaglia}
\email{francesco.sinigaglia@phd.unipd.it}

\author[0000-0002-0639-8043]{Francesco Sinigaglia}
\affiliation{Department of Physics and Astronomy, Università degli Studi di Padova, Vicolo dell’Osservatorio 3, I-35122, Padova, Italy}
\affiliation{INAF - Osservatorio Astronomico di Padova, Vicolo dell’Osservatorio 5, I-35122, Padova, Italy}
\affiliation{Instituto de Astrof\'isica de Canarias, Calle via L\'actea s/n, E-38205, La  Laguna, Tenerife, Spain}
\affiliation{Departamento  de  Astrof\'isica, Universidad de La Laguna,  E-38206, La Laguna, Tenerife, Spain}

\author[0000-0002-9415-2296]{Giulia Rodighiero}
\affiliation{Department of Physics and Astronomy, Università degli Studi di Padova, Vicolo dell’Osservatorio 3, I-35122, Padova, Italy}
\affiliation{INAF - Osservatorio Astronomico di Padova, Vicolo dell’Osservatorio 5, I-35122, Padova, Italy}

\author[0000-0001-9359-0713]{Ed Elson}
\affiliation{Department of Physics and Astronomy, University of the Western Cape, Robert Sobukwe Rd, 7535 Bellville, Cape Town, South Africa}

\author[0000-0002-6748-0577 ]{Mattia Vaccari}
\affiliation{Department of Physics and Astronomy, University of the Western Cape, Robert Sobukwe Rd, 7535 Bellville, Cape Town, South Africa}
\affiliation{The Inter-University Institute for Data Intensive Astronomy (IDIA), Department of Astronomy, University of Cape Town, Private Bag X3, Rondebosch, 7701, South Africa}
\affiliation{The Inter-university Institute for Data Intensive Astronomy (IDIA), Department of Physics and Astronomy, University of the Western Cape, 7535 Bellville, Cape Town, South Africa}
\affiliation{INAF - Istituto di Radioastronomia, via Gobetti 101, I-40129, Bologna, Italy}

\author[0000-0001-8312-5260]{Natasha Maddox}
\affiliation{University Observatory, Faculty of Physics, Ludwig-Maximilians-Universit\"at, Scheinerstr. 1, 81679 Munich, Germany}

\author[0000-0003-3599-1521]{Bradley S. Frank}
\affiliation{South African Radio Astronomy Observatory, 2 Fir Street, Black River Park, Observatory, 7925, South Africa}
\affiliation{The Inter-University Institute for Data Intensive Astronomy (IDIA), Department of Astronomy, University of Cape Town, Private Bag X3, Rondebosch, 7701, South Africa}
\affiliation{The Inter-university Institute for Data Intensive Astronomy (IDIA), Department of Physics and Astronomy, University of the Western Cape, 7535 Bellville, Cape Town, South Africa}
\affiliation{Department of Astronomy, University of Cape Town, Private Bag X3, Rondebosch 7701, South Africa}

\author[0000-0001-7039-9078]{Matt J. Jarvis}
\affiliation{Oxford Astrophysics, Denys Wilkinson Building, University of Oxford, Keble Rd, Oxford, OX1 3RH, UK}
\affiliation{Department of Physics and Astronomy, University of the Western Cape, Robert Sobukwe Rd, 7535 Bellville, Cape Town, South Africa}

\author[0000-0002-0616-6971]{Tom Oosterloo}
\affiliation{Kapteyn Astronomical Institute, University of Groningen, Landleven 12, NL-9747 AD Groningen, the Netherlands}
\affiliation{ASTRON, Netherlands Institute for Radio Astronomy, Oude Hoogeveensedijk 4, NL-7991 PD Dwingeloo, the Netherlands}

\author[0000-0003-2842-9434]{Romeel Davé}
\affiliation{SUPA, Institute for Astronomy, University of Edinburgh, Royal Observatory, Edinburgh EH9 3HJ, UK}
\affiliation{Department of Physics and Astronomy, University of the Western Cape, Robert Sobukwe Rd, 7535 Bellville, Cape Town, South Africa}
\affiliation{South African Astronomical Observatories, Observatory, Cape Town 7925, South Africa}

\author[0000-0001-7116-9303]{Mara Salvato}
\affiliation{Max Planck Institute for Extraterrestrial Physics, Giessembachstrasse 1, D-857498, Garching}

\author[0000-0002-3930-2757]{Maarten Baes}
\affiliation{Sterrenkundig Observatorium, Universiteit Gent, Krijgslaan 281 S9, 9000 Gent, Belgium}

\author[0000-0003-4169-9738]{Sabine Bellstedt}
\affiliation{International Centre for Radio Astronomy Research (ICRAR), University of Western Australia, 35 Stirling Highway, Crawley, WA 6009, Australia}

\author[0000-0003-0492-4924]{Laura Bisigello}
\affiliation{Department of Physics and Astronomy, Università degli Studi di Padova, Vicolo dell’Osservatorio 3, I-35122, Padova, Italy}
\affiliation{INAF - Osservatorio Astronomico di Padova, Vicolo dell’Osservatorio 5, I-35122, Padova, Italy}

\author[0000-0002-2326-7432]{Jordan D. Collier}
\affiliation{The Inter-University Institute for Data Intensive Astronomy (IDIA), Department of Astronomy, University of Cape Town, Private Bag X3, Rondebosch, 7701, South Africa}
\affiliation{The Inter-university Institute for Data Intensive Astronomy (IDIA), Department of Physics and Astronomy, University of the Western Cape, 7535 Bellville, Cape Town, South Africa}
\affiliation{School of Science, Western Sydney University, Locked Bag 1797, Penrith, NSW 2751, Australia}
\affiliation{CSIRO Astronomy and Space Science, PO Box 1130, Bentley, WA, 6102, Australia}

\author[0000-0002-3238-8359]{Robin H. W. Cook}
\affiliation{International Centre for Radio Astronomy Research (ICRAR), University of Western Australia, 35 Stirling Highway, Crawley, WA 6009, Australia}

\author[0000-0003-3085-0922]{Luke J. M. Davies}
\affiliation{International Centre for Radio Astronomy Research (ICRAR), University of Western Australia, 35 Stirling Highway, Crawley, WA 6009, Australia}

\author[0000-0002-6149-0846]{Jacinta Delhaize}
\affiliation{Department of Astronomy, University of Cape Town, Private Bag X3, Rondebosch 7701, South Africa}

\author[0000-0001-9491-7327]{Simon P. Driver}
\affiliation{International Centre for Radio Astronomy Research (ICRAR), University of Western Australia, 35 Stirling Highway, Crawley, WA 6009, Australia}

\author[0000-0003-0247-1204]{Caroline Foster}
\affiliation{School of Physics, University of New South Wales, Sydney, NSW 2052, Australia}

\author[0000-0001-6615-5492]{Sushma Kurapati}
\affiliation{Department of Astronomy, University of Cape Town, Private Bag X3, Rondebosch 7701, South Africa}

\author[0000-0003-3021-8564]{Claudia del P. Lagos}
\affiliation{International Centre for Radio Astronomy Research (ICRAR), University of Western Australia, 35 Stirling Highway, Crawley, WA 6009, Australia}
\affiliation{ARC Centre of Excellence for All Sky Astrophysics in 3 Dimensions (ASTRO 3D), Australia}

\author[0000-0003-1731-0497]{Christopher Lidman}
\affiliation{Centre for Gravitational Astrophysics, College of Science, The Australian National University, ACT 2601, Australia}
\affiliation{The Research School of Astronomy and Astrophysics, The Australian National University, ACT 2601, Australia}

\author[0000-0001-5175-939X]{Pavel E. Mancera Piña}
\affiliation{Kapteyn Astronomical Institute, University of Groningen, Landleven 12, NL-9747 AD Groningen, the Netherlands}
\affiliation{ASTRON, Netherlands Institute for Radio Astronomy, Oude Hoogeveensedijk 4, NL-7991 PD Dwingeloo, the Netherlands}

\author[0000-0002-2838-3010]{Martin J. Meyer}
\affiliation{International Centre for Radio Astronomy Research (ICRAR), University of Western Australia, 35 Stirling Highway, Crawley, WA 6009, Australia}
\affiliation{ARC Centre of Excellence for All Sky Astrophysics in 3 Dimensions (ASTRO 3D), Australia}

\author[0000-0002-5136-7983]{K. Moses Mogotsi}
\affiliation{South African Astronomical Observatory, P.O. Box 9, Observatory, 7935, Cape Town, South Africa}
\affiliation{Southern African Large Telescope, P.O. Box 9, Observatory, 7935, Cape Town, South Africa}

\author[0000-0002-9160-391X]{Hengxing Pan}
\affiliation{Department of Physics and Astronomy, University of the Western Cape, Robert Sobukwe Rd, 7535 Bellville, Cape Town, South Africa}
\affiliation{Oxford Astrophysics, Denys Wilkinson Building, University of Oxford, Keble Rd, Oxford, OX1 3RH, UK}

\author[0000-0003-4100-0173; ]{Anastasia A. Ponomareva}
\affiliation{Oxford Astrophysics, Denys Wilkinson Building, University of Oxford, Keble Rd, Oxford, OX1 3RH, UK}

\author[0000-0001-9680-7092]{Isabella Prandoni}
\affiliation{INAF - Istituto di Radioastronomia, via Gobetti 101, I-40129, Bologna, Italy}

\author[0000-0001-8461-4701]{Sambatriniaina H. A. Rajohnson}
\affiliation{Department of Astronomy, University of Cape Town, Private Bag X3, Rondebosch 7701, South Africa}

\author[0000-0003-0429-3579]{Aaron S. G. Robotham}
\affiliation{International Centre for Radio Astronomy Research (ICRAR), University of Western Australia, 35 Stirling Highway, Crawley, WA 6009, Australia}

\author[0000-0003-3892-3073]{Mario G. Santos}
\affiliation{Department of Physics and Astronomy, University of the Western Cape, Robert Sobukwe Rd, 7535 Bellville, Cape Town, South Africa}
\affiliation{South African Radio Astronomy Observatory, 2 Fir Street, Black River Park, Observatory, 7925, South Africa}

\author[0000-0002-8418-9001]{Srikrishna Sekhar}
\affiliation{The Inter-University Institute for Data Intensive Astronomy (IDIA), Department of Astronomy, University of Cape Town, Private Bag X3, Rondebosch, 7701, South Africa}
\affiliation{The Inter-university Institute for Data Intensive Astronomy (IDIA), Department of Physics and Astronomy, University of the Western Cape, 7535 Bellville, Cape Town, South Africa}
\affiliation{National Radio Astronomy Observatory, 1003 Lopezville Road, Socorro, NM 87801, USA}
\affiliation{Department of Physics and Astronomy, University of the Western Cape, Robert Sobukwe Rd, 7535 Bellville, Cape Town, South Africa}

\author[0000-0002-0956-7949]{Kristine Spekkens}
\affiliation{Department of Physics and Space Science, Royal Military College of Canada, PO Box 17000, Station Forces, Kingston, Ontario, K7K 7B4, Canada}

\author[0000-0002-7921-0785]{Jessica E. Thorne}
\affiliation{International Centre for Radio Astronomy Research (ICRAR), University of Western Australia, 35 Stirling Highway, Crawley, WA 6009, Australia}

\author[0000-0002-9316-763X]{Jan M. van der Hulst}
\affiliation{Kapteyn Astronomical Institute, University of Groningen, Landleven 12, NL-9747 AD Groningen, the Netherlands}

\author[0000-0003-4264-3509]{O. Ivy Wong}
\affiliation{CSIRO Astronomy and Space Science, PO Box 1130, Bentley, WA, 6102, Australia}
\affiliation{International Centre for Radio Astronomy Research (ICRAR), University of Western Australia, 35 Stirling Highway, Crawley, WA 6009, Australia}



\vspace{1cm}

\begin{abstract}

We present the first measurements of \HI{} galaxy scaling relations from a blind survey at $z>0.15$. We perform spectral stacking of 9023 spectra of star-forming galaxies undetected in \HI{} at $0.23<z<0.49$, extracted from MIGHTEE-\HI{} Early Science datacubes, acquired with the MeerKAT radio telescope. We stack galaxies in bins of galaxy properties ($M_*$, SFR, and sSFR, with ${\rm sSFR}\equiv M_*/{\rm SFR}$),
obtaining $\gtrsim 5\sigma$ detections in most cases, the strongest \HI{}-stacking detections to date in this redshift range. With these detections, we are able to measure scaling relations in the probed redshift interval, finding evidence for a moderate evolution from the median redshift of our sample $z_{\rm med}\sim 0.37$ to $z\sim 0$. In particular, low-$M_*$ galaxies ($\log_{10}(M_*/{\rm M_\odot})\sim 9$) experience a strong \HI{} depletion ($\sim 0.5$ dex in $\log_{10}(M_{\rm HI}/{\rm M}_\odot)$), while massive galaxies ($\log_{10}(M_*/{\rm M_\odot})\sim 11$) keep their \HI{} mass nearly unchanged. When looking at the star formation activity, highly star-forming galaxies evolve significantly in $M_{\rm HI}$ ($f_{\rm HI}$, where $f_{\rm HI}\equiv M_{\rm}/M_*$) at fixed SFR (sSFR), while at the lowest probed SFR (sSFR) the scaling relations show no evolution. These findings suggest a scenario in which low-$M_*$ galaxies have experienced a strong \HI{} depletion during the last $\sim4$ Gyr, while massive galaxies have undergone a significant \HI{} replenishment through some accretion mechanism, possibly minor mergers. 
Interestingly, our results are in good agreement with the predictions of the \textsc{simba} simulation. We conclude that this work sets novel important observational constraints on galaxy scaling relations.

\end{abstract}



\section{Introduction} \label{sec:intro}

In the modern paradigm of galaxy formation and evolution, the baryon cycle of galaxies can be investigated through the parametrization of scaling relations linking their physical properties at different cosmic times. 

In this context, the neutral atomic hydrogen (\HI{}) constitutes the fundamental component for H$_2$ assembly and therefore represents the raw fuel for star formation. 
Unveiling \HI{} scaling relations in galaxies is thus a task of paramount importance to understand how the availability of fresh cold gas regulates star formation and galaxy evolution.  

Tight relations at $z\sim 0$ between the \HI{} content of star-forming galaxies and their stellar mass \citep[$M_*$,][]{Huang2012,Maddox2015}, 
star formation rate \citep[SFR,][]{Feldmann2020}, 
and disc size \citep[e.g.][and references therein]{Wang2016} 
among others, have been revealed by large-scale \HI{} galaxy surveys, such as the \HI Parkes All-Sky Survey \citep[HIPASS, ][]{Barnes2001}, 
the Arecibo Legacy Fast ALFA Survey \citep[ALFALFA, ][]{Giovanelli2005} and the GALEX Arecibo SDSS Survey \citep[GASS, ][]{Catinella2010}.

Nonetheless, the \HI{} content of galaxies has been so far included in the global scaling relation picture only at very low redshift ($z<0.15$), due to the intrinsic faintness of the 21-cm hyperfine transition emission line and, hitherto, the limited sensitivity of radio telescopes.

A few blind deep observational efforts --- e.g. the Blind Ultra-Deep HI Environmental Survey \citep[BUDHIES, ][]{Verheijen2007}, 
the COSMOS HI Large Extragalactic Survey \citep[CHILES, ][]{Hess2019} and the Arecibo Ultra-Deep Survey \citep[AUDS, ][]{Hoppman2015} --- have reported sparse detections at $z>0.15$. However, their samples are not large enough to constrain scaling relations. 

In the build-up towards the the Square Kilometre Array (SKA), spectral line stacking \citep[e.g.][]{Zwaan2000} can be adopted as an alternative cheaper observational technique to direct detection, performing an average \HI{} mass ($M_{\rm HI}$) detection of a given galaxy sample. Stacking has been proved to be a powerful tool to investigate many different aspects of galaxy evolution, among which the presence and abundance of \HI{} in galaxy clusters \citep[e.g.][]{Lah2009,Healy2021}, 
scaling relations \citep[e.g.][]{,Brown2017}, 
the $M_{\rm HI}$ content of AGN host galaxies \citep{Gereb2015}, 
and the redshift evolution of the \HI{} cosmic density parameter $\Omega_{\rm HI}$  \citep[e.g.][and references therein]{Delhaize2013,Rhee2013,Chowdhury2020}. 
In particular, \cite{Rhee2013,Rhee2016,Rhee2018} reported tentative detections ($\lesssim 3\sigma$) at $z\sim 0.2$, $z\sim 0.37$ and $z\sim 0.32$, respectively. \cite{Chowdhury2020,Chowdhury2021} claimed $\sim4.5\sigma$ and $\sim 5\sigma$ detections with stacking at $z\sim1$ and $z\sim 1.3$, respectively. 

In this Letter, we perform spectral stacking on MIGHTEE-\HI{} (see \S\ref{sec:mightee} for a description of the MIGHTEE survey) datacubes to study \HI{} scaling relations out to $z\sim0.5$ and report the first direct measurement of such relations at $z>0.15$. In particular, we focus here on the relations between stellar mass, \HI{} mass, and star formation activity, to study how such relations linking key galaxy parameters evolve at $z<0.5$. Moreover, we compare our stacked results with simulated results from the \textsc{simba} cosmological simulation \citep[][]{Dave2019}.  


The paper is structured as follows. In \S\ref{sec:mightee} we introduce the MIGHTEE-\HI{} survey and our \HI{} data. In \S\ref{sec:sample} we present our galaxy sample. In \S\ref{sec:stacking} we summarize the basic principles of the stacking procedure we adopt throughout the paper. \S\ref{sec:results} presents the main results of this work and our interpretation of them. We conclude in \S\ref{sec:conclusions}.

We assume a spatially-flat ($\Omega_{\rm k}=0$) $\Lambda$CDM Cosmology, with cosmological parameters from the latest Planck collaboration results \citep{Planck2018}, i.e. $H_0=67.4$ km ${\rm s}^{-1}$ Mpc$^{-1}$, $\Omega_{\rm m}=0.315$, and $\Omega_\Lambda=0.685$.


\section{\HI{} data from MIGHTEE} \label{sec:mightee}

The MeerKAT International GigaHertz Tiered Extragalactic Exploration Large Survey Program \citep[MIGHTEE,][]{Jarvis2016} is a survey, conducted with the MeerKAT radio interferometer \citep[e.g. ][]{Jonas2016}
, observing four deep, extragalactic fields (COSMOS, XMMLSS, ECDFS, ELAIS-S), characterized by a wealth of multi-wavelength data made available by past and ongoing observational efforts. 
MeerKAT is the SKA precursor located in South Africa and comprises 64 offset Gregorian dishes ($13.5$ m diameter main reflector and $3.8$ m sub-reflector), equipped with receivers in UHF–band ($580< \nu <1015$ MHz), L–band ($900< \nu <1670$ MHz), and S–band ($1750< \nu <3500$ MHz).

The MeerKAT data were acquired in spectral and full Stokes mode, thereby making MIGHTEE a spectral line, continuum and polarization survey. In this paper, we make use of the Early Science \HI{} spectral line data from MIGHTEE, presented in  \citet[][]{Maddox2021}; Frank et al. in (prep.).
The observations were conducted between mid-2018 and mid-2019 and targeted $\sim3.5$ deg$^2$ in the XMMLSS field and $\sim1.5$ deg$^2$ in the COSMOS field. These observations were performed with the full array (64 dishes) in L–band, using the 4k correlator mode ($209\,\rm{kHz}$, corresponding to $52$ km s$^{-1}$ at $z=0.23$ and $56$ km s$^{-1}$ at $z=0.49$). The MIGHTEE–\HI{} Early Science visibilities were processed with the \texttt{ProcessMeerKAT} calibration pipeline. The pipeline is \texttt{Casa}-based and performs standard calibration routines and strategies for the spectral line data such as flagging, bandpass and complex gain calibration. The continuum subtraction was done in both the visibility and image domain using standard \texttt{Casa} routines \textsc{uvsub} and \textsc{uvcontsub}. Residual visibilities after continuum subtraction were imaged using \texttt{Casa}’s task \textsc{tclean} with Briggs' weighting (\textsc{robust}=0.5). Eventually, median filtering was applied to the resulting datacubes to reduce the impact of errors due to continuum subtraction. A full description of the data reduction strategy and data quality assessment will be presented in Frank et al. (in prep).

\begin{table}
    \centering
    \begin{tabular}{ll}
    \toprule
    \toprule
    \textbf{MIGHTEE-\HI{} data} & \\
    \toprule
    \textbf{Survey parameter} & \textbf{Value}\\
    \midrule
    Field & COSMOS \\
    Area     & $1.5$ deg$^2$ \\
    Integration time & $16$h \\
    Frequency resolution & $209$ kHz\\
    Velocity resolution & $52$ km s$^{-1}$ at $z=0.23$\\
    Frequency range & $0.950-1.050$ GHz \\
    Velocity range & $68952-146898$ km s$^{-1}$ \\
    Beam (FWHM) &  $14.5^{\prime\prime}\times11.0^{\prime\prime}$ \\
    \bottomrule
    \bottomrule
    \end{tabular}
    \caption{Summary of the details of MIGHTEE-\HI{} data presented in \S\ref{sec:mightee} and used in this paper.}
    \label{tab:mightee}
\end{table}

Out of the full MIGHTEE-\HI{} dataset, we make use of MIGHTEE-\HI{} datacubes covering the COSMOS field. Our analysis is limited to the redshift interval $0.23<z<0.49$. At these redshifts, MIGHTEE-\HI{} data are found to have well-behaved Gaussian noise, with median \HI{} noise rms increasing with decreasing frequency, from $85$ $\mu$Jy beam$^{-1}$ at $\nu\sim 1050$ MHz to $135$ $\mu$Jy beam$^{-1}$ at $\nu\sim 950$ MHz (Frank et al., in prep.). We exclude the spectral bands at $0.09<z<0.23$ and $z>0.49$ from our analysis, as they are characterized by strong RFI features \citep[][Frank et al. in prep.]{Maddox2021}. A first preliminary unguided visual source finding reported no direct \HI{} detections at $z>0.23$. A summary of the features of the MIGHTEE-\HI{} data used in this work is provided in Table \ref{tab:mightee}.


\begin{figure*}
    \centering
    \includegraphics[width=\textwidth]{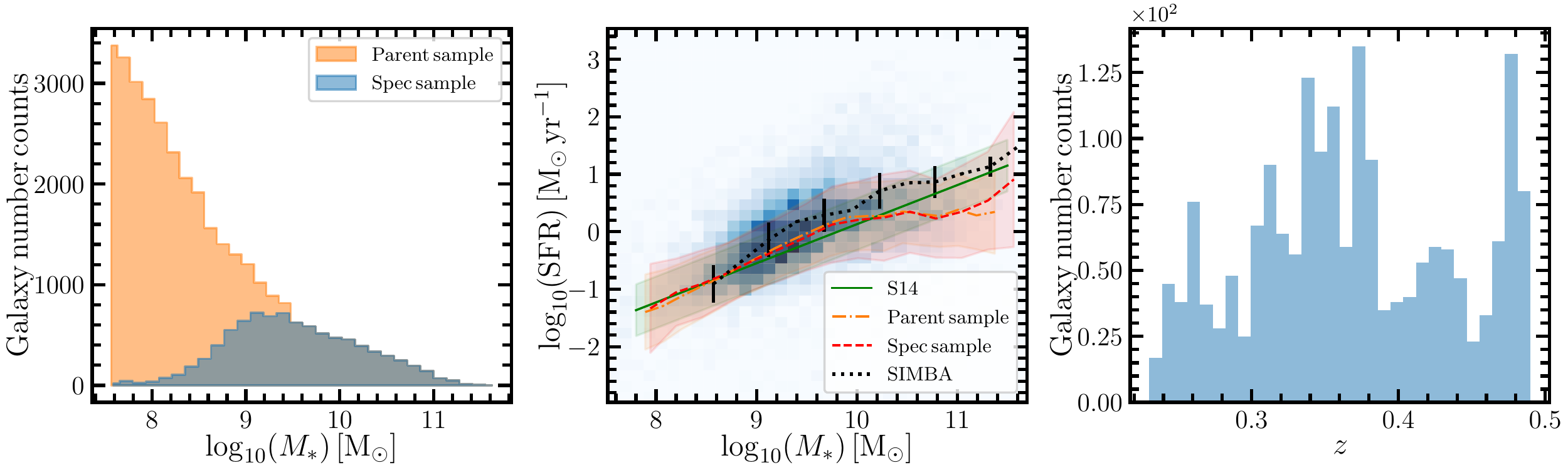}
    \caption{Left: $\log_{10}(M_*)$ distribution of our spectroscopic galaxy sample (blue) and of the parent photometric sample (orange). Center: distribution of our sample (blue colormap histogram) on the $\log_{10}({\rm SFR})-\log_{10}(M_*)$ plane, with Main Sequence parametrization from \cite{Speagle2014}, and average trends from our sample, the parent photometric sample, and the \textsc{simba} simulation overplotted as green solid, red dashed, orange dashed-dotted, and black dotted lines, respectively. Red, orange, and green shaded regions indicate uncertainties on the curves of the same colors. Right: redshift distribution of our spectroscopic galaxy sample.}
    \label{fig:sample}
\end{figure*}

\section{Sample selection}\label{sec:sample}

We select star-forming galaxies at redshift $0.23<z<0.49$ in the COSMOS field \citep{Scoville2007}, with spectroscopic redshift information available. 

We start by considering the latest COSMOS photometric sample publicly available as part of the COSMOS2020 data release \citep[][]{Weaver2021}. This dataset includes estimates of the derived galaxy properties (we consider $M_*$ and SFR in this work) obtained through spectral energy density (SED) fitting. Then, we select star-forming galaxies, according to a color-color $NUV-r/r-J$ plane selection. In particular, quiescent galaxies are defined as those having $M_{NUV}-M_r > 3(M_r - M_J) + 1$ and $M_{NUV} - M_r > 3.1$, while the remaining galaxies are flagged as star-forming \citep[see][]{Laigle2016}. This defines our parent sample. 

Out of the parent sample, we extract only galaxies having a spectroscopic counterpart. This is done by cross-matching the parent sample with a list of spectroscopic redshifts assembled by querying publicly-available catalogs from spectroscopic surveys of the COSMOS field (Salvato et al., in prep.) and spectroscopic redshifts acquired by the DEVILS survey \citep[][]{Davies2018}. Because the SED fitting was performed by adopting the photometric redshift as estimates for redshift, we checked for outliers in the photometric redshifts determination\footnote{Outliers are defined as those galaxies having: \\$|z_{\rm spec}-z_{\rm phot}|>0.15\times(1+z_{\rm spec})$.}. We find that the outliers constitute $\lesssim 5\%$ of our sample, and excluded them. 

Moreover, we explicitly cross-checked the accuracy of the SFR determination, which can induce substantial deviations when obtained through SED fitting in the absence of FIR bands data in the case of dust-attenuated galaxies. In particular, where applies, we compare our SFR estimates with the ones obtained relying on independent FIR measurements (not used in the SED fitting mentioned above) presented in \citet{Yin2018}. We find that there are no systematic offsets, and that the two different SFR estimates are in general consistent within a scatter of $\sim 0.4$ dex, comparable to IR-based uncertainties on SFR. On the other hand, weakly- and non-attenuated galaxies should have an accurate SFR estimate due to the richness of photometric bands in the COSMOS2020 catalog, provided that photo-$z$ outliers are excluded. We also checked for the presence of AGNs in our sample by cross-matching it with the radio-selected AGN catalog built by \citet{Smolcic2017}. AGNs constitute a $\sim 3.5\%$ fraction, and we do not explicitly remove them from our sample. Their impact on our stacked results will be addressed in future publications.

The COSMOS photometric sample has been shown to be $90\%$ complete in $M_*$ down to $\log_{10}(M_*{\rm /M_\odot})\sim8.5$ at $z\lesssim 0.5$ \citep[e.g.][]{Laigle2016}. When comparing the $M_*$ distributions (Fig. \ref{fig:sample}, left panel), the spectroscopic sample appears to be faithfully sampling the parent photometric sample at $\log_{10}(M_*/{\rm M_\odot})\gtrsim 9.5$, while the two distributions differ at $\log_{10}(M_*/{\rm M_\odot})< 9.5$. However, one should take into account that our spectroscopic sample is the result of the combination of several different surveys, having different targets and performed with different survey strategies. 
Hence, the resulting incompleteness at small $M_*$ is not surprising. Nonetheless, as long as the galaxies of the spectroscopic sample at a given $M_*$ are representative of the star-forming galaxy population of the photometric sample at the same $M_*$, the impact of the incompleteness of the sample is minimized by the fact that we group galaxies into bins. 

As a cross-check, we compare the spectroscopic and photometric samples in the SFR$-M_*$ plane (Fig. \ref{fig:sample}, central panel). We group galaxies into bins of $M_*$, and plot the trend connecting the average SFR in different bins (red dashed for our sample, orange dashed-dotted for the parent photometric sample), with uncertainty given by the standard deviation, reported as shaded areas. The spectroscopic sample  is in excellent agreement with the photometric sample, and both well span the Main Sequence parametrization provided by \citet{Speagle2014} across its dispersion (above and below the mean relation), at least at $\log_{10}(M_*/{\rm M_\odot})\gtrsim 8.5$. 
Therefore, in the remainder of the paper we assume our sample is not biased by selection effects and include in our analysis all the galaxies down to $\log_{10}(M_*/{\rm M_\odot})\sim 8.5$.
Our final spectroscopic sample consists of $9023$ sources.

\section{Stacking procedure} \label{sec:stacking}

We adopt a standard spectral line stacking procedure throughout the paper \citep[see e.g.][]{Healy2019}. We obtain \HI{} spectra by extracting cubelets around individually-undetected galaxies from the full datacubes with suitable apertures ($3\times$FWHM of the beam on the image plane, $\pm1500$ km s$^{-1}$ along the velocity axis) and integrating them over angular coordinates. We choose these apertures to ensure that the whole flux emitted by galaxies is included in the cubelets. The angular aperture corresponds to $\sim 130$ physical kpc at $z=0.23$ (i.e. minimum aperture), larger than the typical H{\scriptsize I} disk size. This breaks the degeneracy between underestimating and overestimating the flux depending on cubelet aperture, leaving us only with the problem of subtracting the flux contamination by nearby sources. Optical coordinates and spectroscopic redshift measurements are used to define the center of the cubelets. Each spectrum at observed frequency $\nu_{\rm obs}$ is then de-redshifted to its rest-frame frequency $\nu_{\rm rf}$ through $\nu_{\rm rf}=\nu_{\rm obs}(1+z)$ and converted to units of velocity as $v=cz$. Furthermore, spectra are resampled to a reference spectral template, to ensure that all the spectra are binned the same manner in the spectral direction. We convert spectra from units of flux to units of $M_{\rm HI}$ (per velocity channel) \citep[e.g.][]{Zwaan2001}:
\begin{equation*}
\centering 
    M_{\rm HI}(v) = (2.356\times10^5)\,D_{\rm L}^2\, S(v)\,(1+z)^{-1} \, {\rm M}_{\odot}\, {\rm km}^{-1} \, {\rm s} 
\end{equation*}
where $D_{\rm L}$ is the luminosity distance of the galaxy in units Mpc, $S(v)$ is the 21-cm spectral flux density in units Jy and $(1+z)^{-1}$ is a correction factor accounting for the flux reduction due to the expansion of the Universe.
Lastly, we co-add all the spectra together. The stacked spectrum can then be expressed as 
\begin{equation}
    \braket{M_{\rm HI}(v)} = \frac{\sum_{i=0}^{n_{\rm gal}} M_{{\rm HI},i}(v) \times \,w_i \times f_i }{\sum_{i=0}^{n_{\rm gal}} w_i \times f_i^2}
\end{equation}
where $n_{\rm gal}$ is the number of co-added spectra, and $f_i$ and $w_i$ are the average primary beam transmission and the weight assigned to each source. In the standard unweighted case, $w_i=1$ and $\sum_i w_i=n_{\rm gal}$. This equation implements primary beam correction following the procedure detailed in \cite{Gereb2013}. Throughout the paper, we assume the \cite{Fabello2011a} weighting scheme ($w_i=1/\sigma_{\rm{ rms},i}^2$). The 1$\sigma$ noise uncertainty (in units $M_{\rm HI}$) is evaluated by computing the root mean square (rms) of the noisy channels $\sigma_{\rm rms}$ of the stacked spectrum, i.e. those channels outside the spectral interval integrated to compute $M_{\rm HI}$.

To further confirm the legitimacy of our detection, we also generate a reference spectrum obtained by stacking noise spectra (one noise spectrum per galaxy) extracted at randomized positions.
The positions of the noise spectra are obtained by adding a fixed angular offset to the centre of each galaxy cubelet in a random direction, and defined over the same spectral range as the corresponding galaxy cubelet. The angular offset ($100^{\prime\prime}$) is chosen to guarantee that the reference spectrum is extracted close to the galaxy spectrum, although without overlaps. Also, we double-check that the reference spectrum of each galaxy has no overlaps with other known optical galaxies, and reject it and draw a new one if there is any overlap. 

We compute the integrated signal-to-noise ratio of the final stacked spectrum as:
\begin{equation}\label{eq:snr}
{\rm SNR}=\sum_i^{N_{\rm ch}} \braket{S_i} /(\sigma_{\rm rms}\sqrt{N_{\rm ch}}) 
\end{equation}
where $\braket{S_i}$ is the stacked spectrum, and $N_{\rm ch}$ is the number of channels over which the integration is performed \citep[e.g.,][]{Healy2019}. We estimate uncertainties on the stacked spectrum by applying jackknife resampling 
to the galaxy sample, eliminating one galaxy at a time. 

We address the problem of flux contamination due to source confusion using detailed MeerKAT-like simulated datacubes. In particular, we use the \cite{Obreschkow2014} flux-limited mock galaxy catalog, based on the SKA Simulated Skies semi-analytic simulations (S$^3$-SAX) 
to inject galaxies with realistic \HI{} masses and clustering into a blank synthetic datacube matching the same angular and spectral size as our observations \citep[see also][]{Elson2016,Elson2019}.
Then, following \citet{Elson2016}, we decomposed the spectrum extracted for each target galaxy into contributions from the actual target, and contributions from nearby contaminating galaxies. In this way, we could accurately calculate that the level of contamination is $\lesssim 10\%$ in all the studied cases, and subtract a fixed $10\%$ contribution from the output $\braket{M_{\rm HI}}$ from stacking. 



\begin{figure}
    \centering
    \includegraphics[width=\columnwidth]{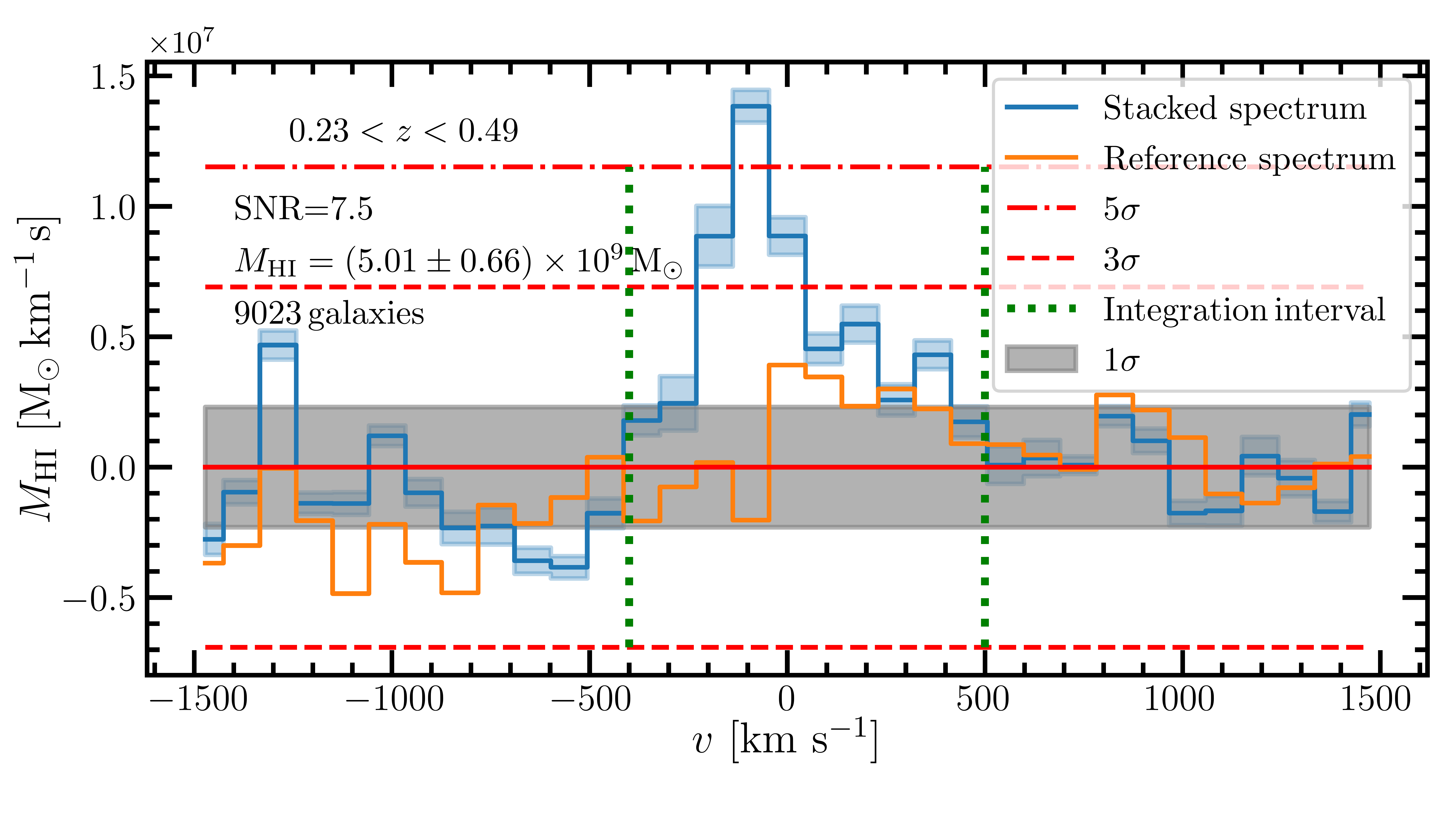}
    \vspace{-0.7cm}
    \caption{Stacked $M_{\rm HI}$ spectra obtained in the full $0.23<z<0.49$ redshift range we investigate in this paper. 
    The blue solid line and the blue shaded region show the stacked spectrum and the associated jackknife uncertainties, respectively. The orange solid line and the gray shaded area indicate the reference spectrum (see text) and $\sigma_{\rm rms}$, respectively. The red dashed and dashed-dotted horizontal lines mark the $3\sigma_{\rm rms}$ and $5\sigma_{\rm rms}$ noise levels, respectively. The green vertical line shows the velocity integration limits in the $M_{\rm HI}$ computation.}
    \label{fig:spectra_zcut}
\end{figure}

\begin{figure*}
    \centering
    \includegraphics[width=\textwidth]{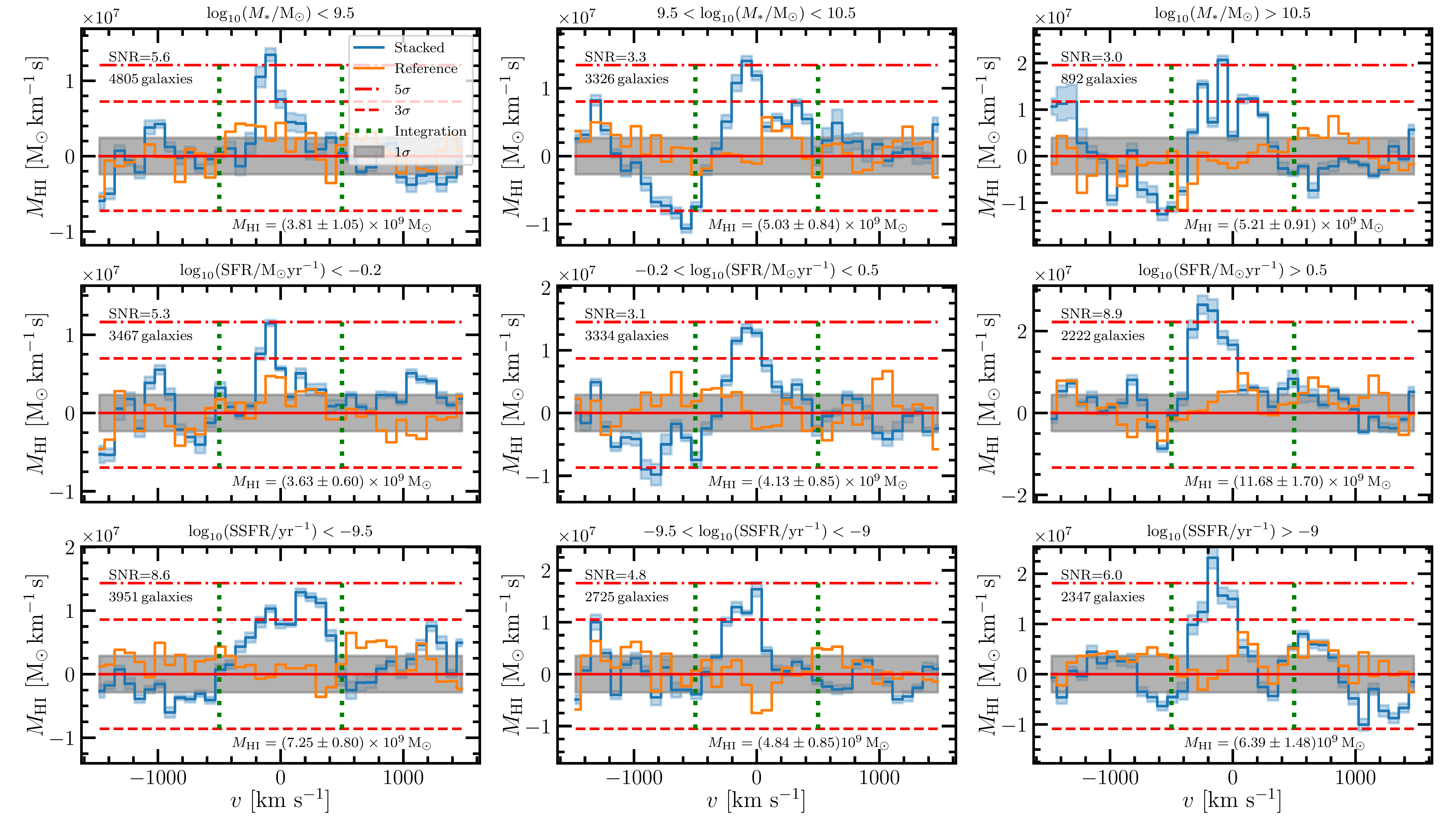}
    \caption{Stacked $M_{\rm HI}$ spectra obtained in different intervals of $M_*$ (top row), SFR (mid row), and sSFR (bottom row). The title of each panel indicates the corresponding interval of galaxy properties. Lines and colors follow the same scheme as Fig. \ref{fig:spectra_zcut}.} 
    \label{fig:spectra_propcut}
\end{figure*}


\section{Results and discussion} \label{sec:results}

In this section we present the main results of this paper. We first discuss the results yielded by stacking in the $0.23<z<0.49$ redshift range, and then compare with known scaling relations at mean redshift $z\sim0$. 

\subsection{Stacking at $0.23<z<0.49$ on MIGHTEE-\HI{} data}

We first produce a global stack using the full galaxy sample. Fig. \ref{fig:spectra_zcut} shows the resulting stacked spectrum. 
As anticipated, we detect signal at $\gtrsim 5\sigma$. 
This represents the strongest detection obtained to date with stacking in this redshift interval. The corresponding $\braket{M_{\rm HI}}$ measurement and integrated SNR are reported in Fig. \ref{fig:spectra_zcut}. 

In Fig. \ref{fig:spectra_propcut} we present the stacked spectra obtained in bins of $M_*$ (top row), SFR (mid row) and sSFR (bottom row). Here, we fix empirically the bin limits to find a compromise between having enough sources per bin to claim a detection, and dissecting the $\log_{10}({\rm SFR})-\log_{10}(M_*)$ in meaningful intervals. 
Our scaling relations are evaluated at the median redshift of our sample $z_{\rm med}\sim 0.37$ (see Fig. \ref{fig:sample}, right panel, for the redshift distribution of our sample). 
To address the potential impact of the fact that the redshift distributions of the subsamples defined in different galaxy properties bins may peak at redshifts higher/lower than $z\sim 0.37$ due to selection effects, we compute the median redshift of each subsample obtained with the aforementioned property cuts. The results we obtain are:
\begin{itemize}
    \item $M_*$ bins: $z_{\rm med}=\{0.36, 0.37, 0.36\}$;
    \item SFR bins: $z_{\rm med}=\{0.35, 0.37, 0.38\}$;
    \item sSFR bins: $z_{\rm med}=\{0.36, 0.36, 0.38\}$
\end{itemize}
We notice that galaxies in all the bins have distribution peaking around the global median redshift $z_{\rm med,tot}\sim 0.37$, with maximum percentage deviation $\Delta z_{\rm max}/z_{\rm med,tot}\sim 5\%$, i.e. a negligible effect. As a result, we regard our results not to be affected by redshift biases.


In this case, we detect signal at $\gtrsim 5\sigma$ in six bins and at $\gtrsim 3\sigma$ in the remaining three bins. The corresponding $\braket{M_{\rm HI}}$ measurements are reported inside the panels in Fig. \ref{fig:spectra_propcut}, together with the resulting integrated SNR and the number of co-added spectra. 
We notice that some stacks present non-negligible negative mass structures at $|v|>500\,\rm{km\,s}^{-1}$, where only random noise should be present. The origin of these features may be due to continuum oversubtraction, or residual RFI. To test the possible effect of RFI, we repeated the stacking procedure after carefully flagging the frequency bands affected by RFI (Frank et al., in prep.) and conservatively excluding the galaxies lying in such regions. This is done to mitigate the possible negative impact of badly behaved spectra, whose importance was not already suppressed by the weighting scheme. After this operation, the resulting stacked spectra no longer present outlier negative mass structures, and the $M_{\rm HI}$ estimates are in excellent agreement with the ones obtained with the full sample, indicating that our results are not biased by this issue.

To fully account for these features, in the three worst cases (central and right panel in the top row, central panel in the central row of Fig. \ref{fig:spectra_propcut}) we include an additional uncertainty term in Eq. \ref{eq:snr}, replacing $\sigma_{\rm rms}$ with $\sigma=\sigma_{\rm rms} + \sigma_{\rm dip}$. We estimate $\sigma_{\rm dip}$ empirically in such  way to reduce the statistical significance of the negative structures below the detection threshold (conservatively set at $2.5\sigma$). The final SNR shown in Fig. \ref{fig:spectra_propcut} is the conservative estimate obtained after this additional operation, while previously the calculation based only on $\sigma_{\rm rms}$ returned SNR$>5$ in all the three cases.

\begin{figure*}
    \centering
    \includegraphics[width=\textwidth]{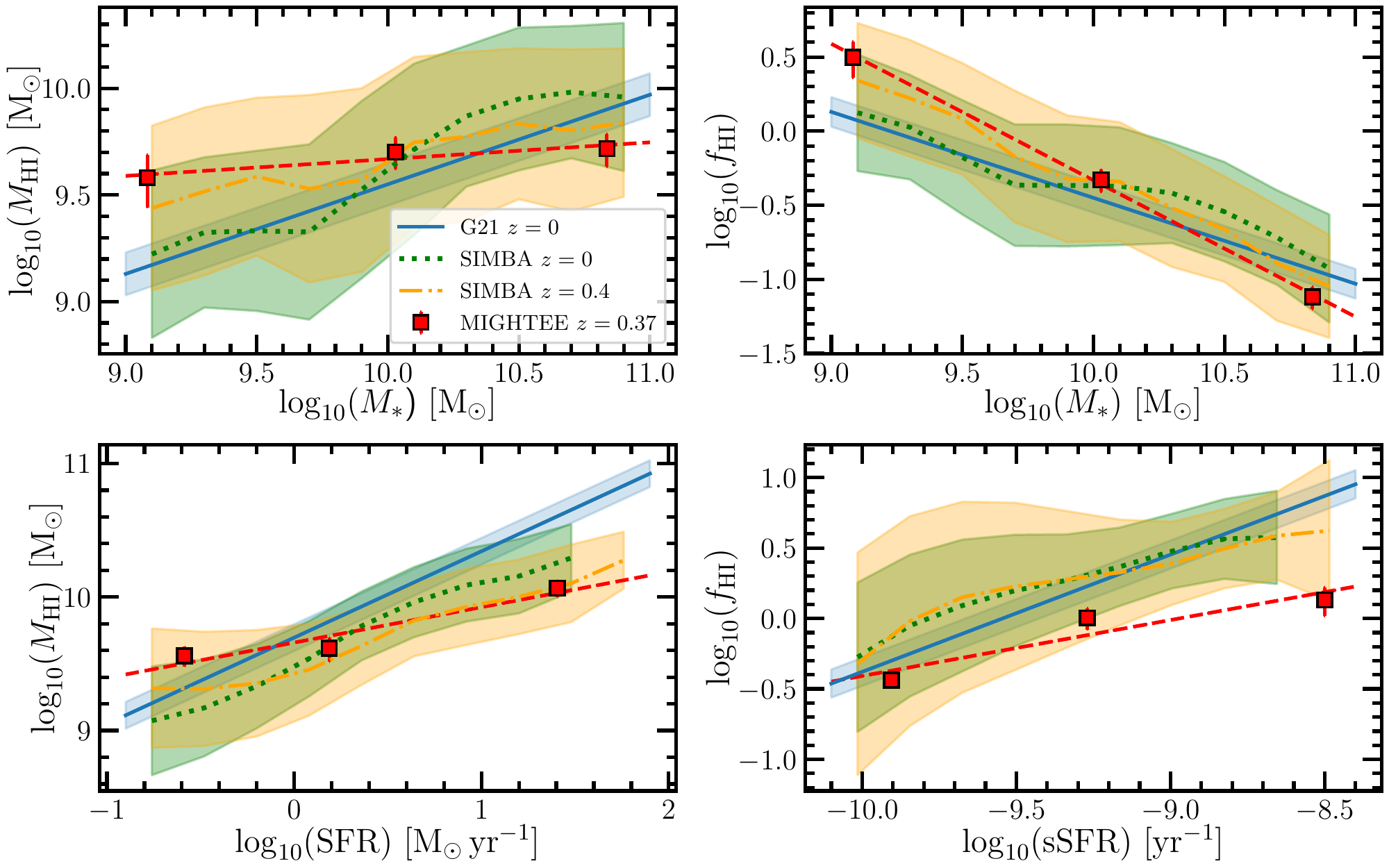}
    \caption{Star-forming galaxy \HI{} scaling relations: $\log_{10}(M_{\rm HI})-\log_{10}(M_*)$ (top left), $\log_{10}(f_{\rm HI})-\log_{10}(M_*)$ (top right), $\log_{10}(M_{\rm HI})-\log_{10}({\rm SFR})$ (bottom left), and $\log_{10}(f_{\rm HI})-\log_{10}({\rm sSFR})$. Our \HI{} stacking results at median redshift $z\sim 0.37$ are displayed as red square dots. Uncertainties on red square dots along the x-axis, not shown for the sake of clarity of visualization, correspond to the width of the bins. The red dashed line represents the fit to our data. Our reference results at $z\sim 0$ \citep[G21,][]{Guo2021} are plotted as a blue solid line. Green dotted and yellow dashed-dotted lines show the predictions of the \textsc{simba} cosmological hydrodynamic simulation at $z=0$ and $z=0.4$, respectively, for comparison. Shaded regions indicate $1\sigma$ uncertainties.}
    \label{fig:scaling_relations}
\end{figure*}

\subsection{Scaling relations at $z\sim 0$}

We adopt as fiducial observational results at $z\sim 0$ the findings presented in \cite{Guo2021} (G21 hereafter). Therein the authors investigate the inter-dependence of $M_{\rm HI}$, $M_*$ and SFR, among others, performing spectral stacking on cross-matched ALFALFA \citep[][]{Haynes2018} and Sloan Digital Sky Survey \citep[SDSS,][]{York2000} data. Results are in close agreement with previous results \citep[e.g.][]{Brown2017,Catinella2018}, where applicable. 
Furthermore, scaling relations are assessed separately for star-forming and quenched galaxies. Therefore, within this work we make comparisons only with the star-forming galaxy results.   

\subsection{\textsc{simba}: reference cosmological hydro simulation}

\textsc{simba} is a cosmological hydrodynamic simulation run with the \texttt{GIZMO} meshless finite mass hydrodynamics, employing $N_{\rm dm}=1024^3$ dark matter particles and $N_{\rm gas}=1024^3$ gas elements in a $V=(100\,{\rm Mpc}\,h^{-1})^3$ comoving volume. The \textsc{simba} fiducial model adopts and updates star formation and feedback sub-grid prescriptions used in the \textsc{mufasa} simulation \citep[][]{Dave2016}, and introduces the treatment of black hole growth and accretion from cold and hot gas. Moreover, models for on-the-fly dust production, growth, and destruction, and \HI{} and H$_2$ abundance computation are implemented  \citep[see][ and references therein for details]{Dave2019}. 

In this work, we make use of \textsc{simba} galaxy catalogs at $z=0$ and $z=0.4$ and select only star-forming galaxies with $\log_{10}(M_*)>8.5$, consistent with our observational data, by separating the star-forming and quenched populations in the $M_*-{\rm SFR}$ plane. In particular, we flag as star-forming those galaxies with $\log_{10}({\rm SFR}/({\rm M}_\odot \, {\rm yr}^{-1}))>\log_{10}(M_*/{\rm M}_\odot)-10.3$ and $\log_{10}({\rm SFR}/({\rm M}_\odot \, {\rm yr}^{-1}))>0.9\times\log_{10}(M_*/{\rm M}_\odot)-9.2$ at $z=0$ and $z=0.4$, respectively. These equations are obtained from a $3\sigma$ cut below the \cite{Speagle2014} parametrizations at the corresponding redshifts,  slightly modified to account for the mismatch between the \textsc{simba} and the observed Main Sequence and to separate well star-forming and quiescent galaxies.

\subsection{Comparison of scaling relations} \label{sec:scaling_relations_comparison} 

Fig. \ref{fig:scaling_relations} shows a comparison between our resulting scaling relations at $0.23<z<0.49$, the findings of G21 at $z\sim 0$ and the theoretical results obtained by the \textsc{simba} simulation at $z=0$ and $z=0.4$. The top left panel shows the $\log_{10}(M_{\rm HI})-\log_{10}(M_*)$ relation, the top right panel shows the $\log_{10}(f_{\rm HI})-\log_{10}(M*)$ relation, the bottom left panel shows the $\log_{10}(M_{\rm HI})-\log_{10}({\rm SFR})$ relation, and the bottom right panel shows the $\log_{10}(f_{\rm HI})-\log_{10}({\rm sSFR})$ relation. The \textsc{simba} curves were generated by breaking the full galaxy sample into bins of $M_*$ (top left panel), or SFR (bottom left panel), or sSFR (bottom right panel), and computing average and standard deviation in each bin.

We fit a linear model $\log_{10}(Y)=\alpha\log_{10}(X) +\beta$ to our data at $z\sim 0.37$.  We compute the mean values for the fitting parameters and the associated uncertainties ($68\%$ confidence interval) through parametric bootstrap resampling of our data, drawing $10000$ samples and fitting the model with a least-squares minimization to each sample. The resulting best-fitting coefficients are:
\begin{itemize}
    \item $\log_{10}(M_{\rm HI}) = \alpha\log_{10}(M_*)+\beta$:\\ $\alpha=0.08^{+0.07}_{-0.07}$ ($\sim 4.8\sigma$ difference from $z\sim0$), $\beta=8.86^{+0.75}_{-0.74}$;
    \item $\log_{10}(f_{\rm HI})=\alpha\log_{10}(M*)+\beta$:\\ $\alpha=-0.92^{+0.07}_{-0.07}$, $\beta=8.88^{+0.72}_{-0.75}$;
    \item $\log_{10}(M_{\rm HI})=\alpha\log_{10}({\rm SFR})+\beta$:\\ $\alpha=0.26^{+0.04}_{-0.04}$, $\beta=9.65^{+0.44}_{-0.43}$;
    \item $\log_{10}(f_{\rm HI})=\alpha\log_{10}({\rm sSFR})+\beta$:\\ $\alpha=0.40^{+0.07}_{-0.07}$, $\beta=3.56^{+0.69}_{-0.68}$. 
\end{itemize}

The $\log_{10}(M_{\rm HI})-\log_{10}(M_*)$ relation (Fig. \ref{fig:scaling_relations}, top left panel) is found to have comparable normalization at the two redshifts at $\log_{10}(M_*/{\rm M}_\odot)\sim 10.3$, but different slopes, the relation at $z=0$ has a steeper slope than at $z\sim 0.37$.
In particular, low-$M_*$ galaxies ($\log_{10}(M_*/{\rm M_\odot})\sim 9$) are $\sim0.5$ dex more HI-rich at $z\sim 0.37$ than at $z\sim 0$, while massive galaxies ($\log_{10}(M_*/{\rm M_\odot})\sim11$) at $z\sim0.37$ and $z\sim0$ converge to similar $M_{\rm HI}$ values within $0.25$ dex (and compatible within uncertainties). The same result can be visualized in terms of \HI{} fraction in the $\log_{10}(f_{\rm HI})-\log_{10}(M_*)$ plane (Fig. \ref{fig:scaling_relations}, top right panel). 

In terms of star formation properties, the SFR displays a weaker correlation with $M_{\rm HI}$ than at $z\sim 0$ (bottom left panel). 
Galaxies having $\log_{10}({\rm SFR}/({\rm M_{\odot} \, yr^{-1}}))\lesssim 0.5$ show no $M_{\rm HI}$ evolution. Instead, galaxies at $z\sim 0$ with $\log_{10}({\rm SFR}/({\rm M_{\odot} \, yr^{-1}})) > 0.5$ feature an excess in $M_{\rm HI}$ of up to $\sim 0.5$ dex in $\log_{10}(M_{\rm HI}$) at  fixed $\log_{10}({\rm SFR}/({\rm M_{\odot} \, yr^{-1}}))\sim 1.4$ with respect to galaxies at $z\sim 0.37$. Lastly, the $f_{\rm HI}-{\rm sSFR}$ relations at $z\sim 0$ and $z\sim 0.37$ converge at $\log_{10}({\rm sSFR}/{\rm yr^{-1}})\sim -10$, while the galaxies at $z\sim 0.37$ at (fixed) larger ${\rm sSFR}$ have systematically smaller $f_{\rm HI}$ than at $z\sim 0$ (up to $\sim 0.8$ dex).

\vspace{0.5cm}

\subsection{Discussion}

Our findings suggest evidence for moderate evolution of scaling relations from $z\sim0.37$ to $z\sim0$. 
Low-$M_*$ galaxies have depleted their \HI{} reservoirs over the last 4 Gyr. The long \HI{} depletion time due to star formation for these galaxies, $\tau_{\rm dep,HI}(\hat{M}_*=\log_{10}(M_*)\sim 9.1)=M_{\rm HI}(\hat{M}_*)/{\rm SFR}(\hat{M}_*)\sim 15 \, {\rm Gyr}$, suggests that star formation depletes only $\Delta M_{\rm HI}\sim 1\times 10^9 \,{\rm M}_\odot$ (i.e. $\sim 0.15$ dex) over the last 4 Gyr. As a result, we argue that star formation alone could not be able to account for the observed reduction in $M_{\rm HI}$ from $z\sim 0.37$ to $z\sim 0$, and that another \HI{} depletion/removal mechanism may be in act in parallel to star formation. On the other hand, massive galaxies have experienced an efficient replenishment of the \HI{} content in their disks, counteracting the \HI{}  depletion due to star formation and feedback processes. In particular, assuming the H$_2$ formation rate and the SFR are in equilibrium, and that the H$_2$ depletion time $\tau_{\rm H_2}$ at $z<0.5$ is of order $0.1<\tau_{\rm H_2}<1$ Gyr \citep[e.g.][and references therein]{Tacconi2018}, there must be an efficient \HI{} accretion mechanism fuelling massive galaxies \citep[see e.g.][]{Sancisi2008}. This is also consistent with the \HI{} depletion time of massive galaxies that we are able to derive in this work: $\tau_{\rm dep,HI}(\hat{M}_*=\log_{10}(M_*)\sim 10.75)=M_{\rm HI}(\hat{M}_*)/{\rm SFR}(\hat{M}_*)\sim 1-1.2 \, {\rm Gyr}$. The nature of the \HI{} growth into galaxies is still debated, with co-planar accretion from cosmic flows being favoured by observations of MgII absorbers in quasars down to $z\sim 0.2$ and statistical arguments \citep[e.g.][and references therein]{Bouche2013,Peng2021}. Theoretical predictions indicate that at $z<0.5$ the main cold gas accretion mode onto galaxies is not cosmological accretion \citep[as suggested for accretion at high-z by results based on observations, e.g.][]{Conselice2013} or the galactic fountain mechanisms \citep[e.g.][]{Fraternali2017}, but mergers \citep[e.g.][]{SanchezAlmeida2014,Padmanabhan2020}. Furthermore, simulations tell that most of \HI{} in the Universe is already contained in galaxy discs at the probed redshifts \citep[e.g. $\sim 97\%$ at $z=0$ in][]{VillaescusaNavarro2018}, thus making mergers a potential efficient \HI{} transfer mechanism. We speculate that minor mergers between low-$M_*$ \HI{}-rich galaxies and massive galaxies could be the mechanism that mainly refill the latter of \HI{}. Intriguingly, \citet{Jackson2022} recently found evidence in observations that minor mergers play a major role in the formation of \HI{}-rich massive disk galaxies at $z\sim0$, supporting our proposed scenario. On the other hand, \citet{DiTeodoro2014} argue that cold gas transfer through minor mergers at $z\sim 0$ is not able to sustain star formation, even under stringent assumptions. This is however not in direct contrast with our findings, as low-$M_*$ galaxies are found to contain much less \HI{} at $z\sim 0$ than at $z\sim 0.37$. In any case, our conclusions do not exclude the scenario in which smooth accretion from the inter-galactic medium is the dominant cold gas accretion onto galaxies. 

The decrease in $M_{\rm HI}$ observed in highly star-forming galaxies with $\log_{10}({\rm SFR}/({\rm M_{\odot} \, yr^{-1}}))> 0.5$ from $z\sim 0.37$ to $z\sim 0$ suggests that \HI{} features a stronger correlation with star formation at the latter redshift.
Interesting insights are offered by the $f_{\rm HI}-{\rm sSFR}$ relation. In fact, making bins in sSFR corresponds to binning the ${\rm SFR}-M_*$ plane (central panel, Fig. \ref{fig:sample}) with bin limits being diagonal lines in the $\log_{10}(\rm SFR)-\log_{10}(M_*)$ plane. In particular, the three bins roughly correspond to galaxies in the lower (and below), central, and upper (and above) part of the Main Sequence, from lower to higher sSFR, respectively. This suggests that the galaxies at fixed ${\rm sSFR}$ lying above the Main Sequence are the ones experiencing a larger increase of their $f_{\rm HI}$ over the last $\sim 5$ Gyr.

To develop a more complete picture on the SFR evolution, we would need to include also the H$_2$ scaling relations in our framework, which goes however beyond the scope of the paper. We leave such a study for future work.

However, we notice that \HI{} replenishment in massive galaxies is not able to supply enough gas to fully sustain star formation and, hence, prevent the observed reduction of SFR. We speculate that the main reason for this could be that fresh \HI{} accretes onto the outer part of the disk, and takes a significant amount of time to migrate towards the region within the optical radius --- where the bulk of star formation takes place --- due to galaxy angular momentum \citep[][]{Peng2021}.


Interestingly, our findings are in good agreement with the predictions of the \textsc{simba} simulation employing the full baryon physics model. 
The only significant discrepancy is found in the $\log_{10}(f_{\rm HI})-\log_{10}({\rm sSFR})$ plane, and is due to the \textsc{simba} Main Sequence at $z= 0.4$ having a systematic positive SFR offset with respect to the the observed Main Sequence (both the \citet{Speagle2014} parametrization and, even more, the observed sample), especially at large $M_*$. The immediate aim of this comparison is to provide a minimum contextualization of our observational results into a theoretical scenario. However, the agreement between theory and observations offers the unique advantage of being able to use \textsc{simba} as a benchmark to study which are the driving processes that determine the observed scaling relations and to compare it more thoroughly to our data to better constrain models. This goes beyond the scope of the paper, and we leave it for future work.
The full \textsc{simba} suite also includes other runs with only partial modelling of feedback and baryon processes. 
\citet{Dave2020} 
find that the most crucial phenomena to be modeled to reproduce the high-mass end of the \HI{} (and H$_2$) mass function are AGN, X-ray, and jet feedback. 
We plan to perform an in-depth investigation of the impact of these aspects on scaling relations 
in forthcoming publications, comparing our findings with other cosmological simulations and semi-analytical models too.

\section{Conclusions} \label{sec:conclusions}

In this paper we have performed stacking of $9023$ 21-cm undetected star-forming galaxy spectra extracted from MIGHTEE-\HI{} datacubes at $0.23<z<0.49$ in the COSMOS field. In particular, we have subdivided the full sample into galaxy properties subsets with the aim of directly measuring for the first time \HI{} scaling relations at a median redshift $z_{\rm med}\sim 0.37$.

We find moderate evolution of the probed scaling relations from $z\sim 0.37$ to $z\sim 0$, with no significant evolution in the $\log_{10}(M_{\rm HI})-\log_{10}(M_*)$ and $\log_{10}(f_{\rm HI})-\log_{10}(M_*)$ relations at $\log_{10}(M_*/{{\rm M}_\odot})\gtrsim 10$, implying the necessity of an efficient \HI{} replenishment mechanism in massive galaxies. The $\log_{10}(M_{\rm HI})-\log_{10}({\rm SFR})$ and $\log_{10}(f_{\rm HI})-\log_{10}({\rm sSFR})$ relations evidence how 
highly star-forming galaxies evolve significantly in SFR (sSFR) at fixed $M_{\rm HI}$ ($f_{\rm HI}$), while the evolution of scaling relations at lower SFR (sSFR) is milder. 
We argue that the aforementioned \HI{} replenishment mechanism is not able to prevent star formation quenching in massive galaxies. We will further investigate these aspects in forthcoming publications.

We argue that future MIGHTEE-\HI{} data beyond the Early Science dataset will allow us to strengthen the statistical significance of the results, as will enlarge the footprint at $0.23<z<0.49$ from the $\sim 1.5$ deg$^2$ of the COSMOS field used in this paper, to $\sim 20$ deg$^2$ of the final data release.

\section*{acknowledgments}
The authors warmly thank the anonymous referee for the constructive review and helpful
comments they offered.
The authors also thank Alvio Renzini for insightful comments and discussions.
F.S. acknowledges the support of the doctoral grant funded by the University of Padova and by the Italian Ministry of Education, University and Research (MIUR). G.R. acknowledges the support from grant PRIN MIUR 2017 - 20173ML3WW$\char`_$001. M.V. acknowledges financial support from the South African Department of Science and Innovation's National Research Foundation under the ISARP RADIOSKY2020 Joint Research Scheme (DSI-NRF Grant Number 113121) and the CSUR HIPPO Project (DSI-NRF Grant Number 121291). N.M. acknowledges support of the LMU Faculty of Physics. H.P. acknowledges support from the South African Radio Astronomy Observatory (SARAO). S.H.A.R. is supported by the South African Research Chairs Initiative of the Department of Science and Technology and National Research Foundation. C.F. is the recipient of an Australian Research Council Future Fellowship (project number FT210100168) funded by the Australian Government. A.A.P. acknowledges support of the STFC consolidated grant ST/S000488/1 and from the Oxford Hintze Centre for Astrophysical Surveys which is funded through generous support from the Hintze Family Charitable Foundation. I.P. acknowledges financial support from the Italian Ministry of Foreign Affairs and International Cooperation (MAECI Grant Number ZA18GR02) and the South African Department of Science and Technology's National Research Foundation (DST-NRF Grant Number 113121) as part of the ISARP RADIOSKY2020 Joint Research Scheme. J.M.v.d.H. acknowledges support from the European Research Council under the European Union’s Seventh Framework Programme (FP/2007-2013) / ERC Grant Agreement nr. 291531 (HIStoryNU). The MeerKAT telescope is operated by the South African Radio Astronomy Observatory, which is a facility of the National Research Foundation, an agency of the Department of Science and Innovation. We acknowledge the use of the ilifu cloud computing facility – \url{www.ilifu.ac.za}, a partnership between the University of Cape Town, the University of the Western Cape, the University of Stellenbosch, Sol Plaatje University, the Cape Peninsula University of Technology and the South African Radio Astronomy Observatory. The Ilifu facility is supported by contributions from the Inter-University Institute for Data Intensive Astronomy (IDIA – a partnership between the University of Cape Town, the University of Pretoria, the University of the Western Cape and the South African Radio Astronomy Observatory), the Computational Biology division at UCT and the Data Intensive Research Initiative of South Africa (DIRISA).

\bibliography{sample631}{}
\bibliographystyle{aasjournal}



\end{document}